\documentclass{article}

\usepackage{arxiv}

\usepackage[utf8]{inputenc} 
\usepackage[T1]{fontenc}    
\usepackage{url}            
\usepackage{booktabs}       
\usepackage{amsfonts}       
\usepackage{nicefrac}       
\usepackage{microtype}      
\usepackage{graphicx}
\usepackage{natbib}
\setcitestyle{authoryear,open={(},close={)}} 
\usepackage{hyperref}
\usepackage{doi}
\usepackage{booktabs}
\usepackage{multirow}
\usepackage{makecell} 
\usepackage{xcolor}
\usepackage{adjustbox}

\title{Implementing engrams from a machine learning perspective: XOR as a basic motif.}

\date{June 5, 2024}	
\author{
    \href{https://orcid.org/0000-0001-7914-8494}{\includegraphics[scale=0.06]{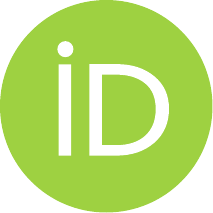}\hspace{1mm}Jesus Marco de Lucas} \\
    \texttt{jesus.marco@csic.es}
    \and
    \href{https://orcid.org/0009-0001-6456-9832}{\includegraphics[scale=0.06]{orcid.pdf}\hspace{1mm}Maria Pe\~na Fern\'andez} \\
    \texttt{penam@ifca.unican.es}
    \and
    \href{https://orcid.org/0000-0002-0157-4765}{\includegraphics[scale=0.06]{orcid.pdf}\hspace{1mm}Lara Lloret Iglesias} \\
    \texttt{lara.cern@gmail.com}
    \\
    Advanced Computing and e-Science Group\\
    Instituto de Física de Cantabria (IFCA) CSIC-Universidad de Cantabria\\
    Santander, ES 39005 SPAIN\\
}

\renewcommand{\headeright}{Short Comment}
\renewcommand{\undertitle}{Short Comment}
\renewcommand{\shorttitle}{\textit{arXiv} Implementing Engrams: XOR Motif}

\hypersetup{
pdftitle={Implementing Engrams: XOR Motif},
pdfsubject={q-bio.NC, q-bio.QM},
pdfauthor={Jesus.~Marco},
pdfkeywords={Engrams, Autoencoders, Prediction by Matching},
}

\begin{document}
\maketitle

\begin{abstract}

We have previously presented the idea of how complex multimodal information could be represented in our brains in a compressed form, following mechanisms similar to those employed in machine learning tools, like autoencoders.
In this short comment note we reflect, mainly with a didactical purpose, upon the basic question for a biological implementation: what could be the mechanism working as a loss function, and how it could be connected to a neuronal network providing the required feedback to build a simple training configuration. 
We present our initial ideas based on a basic motif that implements an XOR switch, using few excitatory and inhibitory neurons. Such motif is guided by a principle of homeostasis, and it implements a loss function that could provide feedback to other neuronal structures, establishing a control system. We analyse the presence of this XOR motif in the connectome of C.Elegans, and indicate the relationship with the well-known lateral inhibition motif. 
We then explore how to build a basic biological neuronal structure with learning capacity integrating this XOR motif. Guided by the computational analogy, we show an initial example that indicates the feasibility of this approach, applied to learning binary sequences, like it is the case for simple melodies.
In summary, we provide didactical examples exploring the parallelism between biological and computational learning mechanisms, identifying basic motifs and training procedures, and how an engram encoding a melody could be built using a simple recurrent network involving both excitatory and inhibitory neurons.  

\end{abstract}


\section{Introduction}

Machine learning solutions used in artificial intelligence open a large pool of interesting questions to compare these computational silicon-based solutions, with biological neuronal systems. 
In a previous paper  (Marco de Lucas, 2023) we proposed the idea that learning in biological systems could involve the compression of information, following an analogy with the use of autoencoders in machine learning. This compression of information into a latent space could be used efficiently to contrast predictions, and also to establish concept cells. Machine learning provides a clear framework to implement this compression of information using an artificial neural network, minimizing a loss function, that computes the difference for each data between the original information that is to be encoded, and the one recovered, or decoded, from the latent space. This loss function is usually minimized using mathematical methods like the gradient descent, that modify the weights connecting the neural network nodes, until such minimum is found. The neural network architecture together with the learned weights provide the compression mechanism, that is able to define the concept nodes over the latent space.
So a key question is what drives learning in a biological system, in comparison to what we use in computational systems. 
First of all, it must be stated that for very simple organisms, the answer could be simply an evolutionary one: those that get a competitive advantage to survive and reproduce will implement, i.e. learn, such mechanisms in their genes, and correspondingly in their neuronal connections. 
However, we already know that the plasticity in all known neural systems goes further, and that we can learn from observations, in a similar way to machine learning.
An initial question is obvious: all machine learning techniques use a “loss function” that is minimized, so we reduce our prediction error. This function is however a mathematical function, computed using real numbers. What could be the biological counterpart, and how does our brain know that its minimization is the right direction to follow? Or in machine learning terms, what drives the credit assignment, i.e. how our neurons know what interconnections must be reinforced or weakened?
In this paper we propose that the basic mechanism could derive from a “homeostatic” rule: we learn by setting up a neural network that is able to handle incoming signals and stabilize the corresponding output, by reinforcing adequate synaptic connections. 
We introduce a simple circuit able to implement this rule: a recurrent network architecture including a biological eXclusive-OR (XOR) switch that establishes the comparison between an incoming signal and the “learned” signal, so providing an effective “loss function”, or minimization mechanism.
We have tested the feasibility of such a circuit using realistic components of biological neuron circuits in the C.Elegans worm, that has been used to model examples of sensory and motor actions (Hasani 2017).
We also propose how such XOR circuit could be integrated in a basic neuronal network, with very few nodes, completing a basic learning block in a RNN (recurrent neural network) scheme, implementing plasticity through the reinforcement of synaptic connections by repeated activation.

\section{Implementing an XOR motif using excitatory and inhibitory neurons}

The implementation of an eXclusive OR (XOR) function is a well-known topic in the study of computational neural networks. 
Our interest in this switch is based on the fact that it could be at the same time a comparator to define a loss function, and a key component to provide an homeostatic control to the propagation of an incoming signal through a biological neural network. The idea is that the switch is used to compare an incoming signal with a reference one implemented by a learning/memory circuit, providing a non-null feedback until both signals are equal, i.e. until the incoming signal is learned/recognized. As soon as this convergence happens, the global circuit does not further propagate the incoming signal, so the homeostatic equilibrium is maintained, but it can also be seen as a minimization of the difference between both signals.
The scheme of this simple XOR circuit is represented in figure 1. It includes two excitatory neurons, labelled as S-1 and S-2, receiving the reference signals, that are connected to another two excitatory neurons, E-1 and E-2, that in turn connect to another neuron, XOR, providing the final result. But the key component is an inhibitory neuron, INH, that receives the input from the two input neurons, and it is able to inhibit the other two excitatory neurons. 
Using adequate weights in the connections among the excitatory neurons, and the inhibitory neuron, this circuit implements the XOR function, providing a 0 when both incoming signals are equal, and a 1 otherwise, provided a discrimination voltage threshold is fixed.

\begin{figure}[h!]
	\centering
       \adjustbox{cfbox=black 1pt}{ 
       \includegraphics[width=0.6\textwidth]{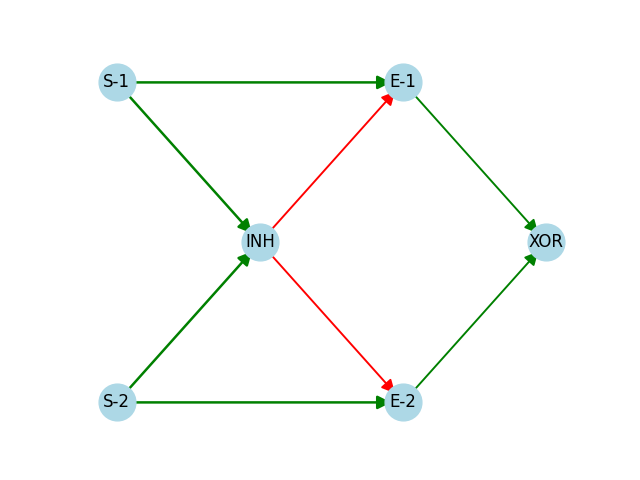} }
	\caption{Scheme of the XOR motif implemented using neurons. Arrows indicate the signal flow through excitatory (green) or inhibitory (red) synaptic connections}
	\label{fig:xor_motif}
\end{figure}

We have tested this dynamic XOR switch using a model for the C. Elegans neurons implemented in SIMULINK, described in (Hasani 2017), where the details of this realistic simulation of these non-spiking neurons are provided. We only use chemical synaptic connections, either excitatory or inhibitory. A key point is the value of the strengths of these synaptic connections, as they will regulate the behaviour of the circuit, that we have heuristically fixed to provide the desired XOR output.

\begin{figure}[h!]
	\centering
       \adjustbox{cfbox=black 1pt}{ 
       \includegraphics[width=0.9\textwidth]{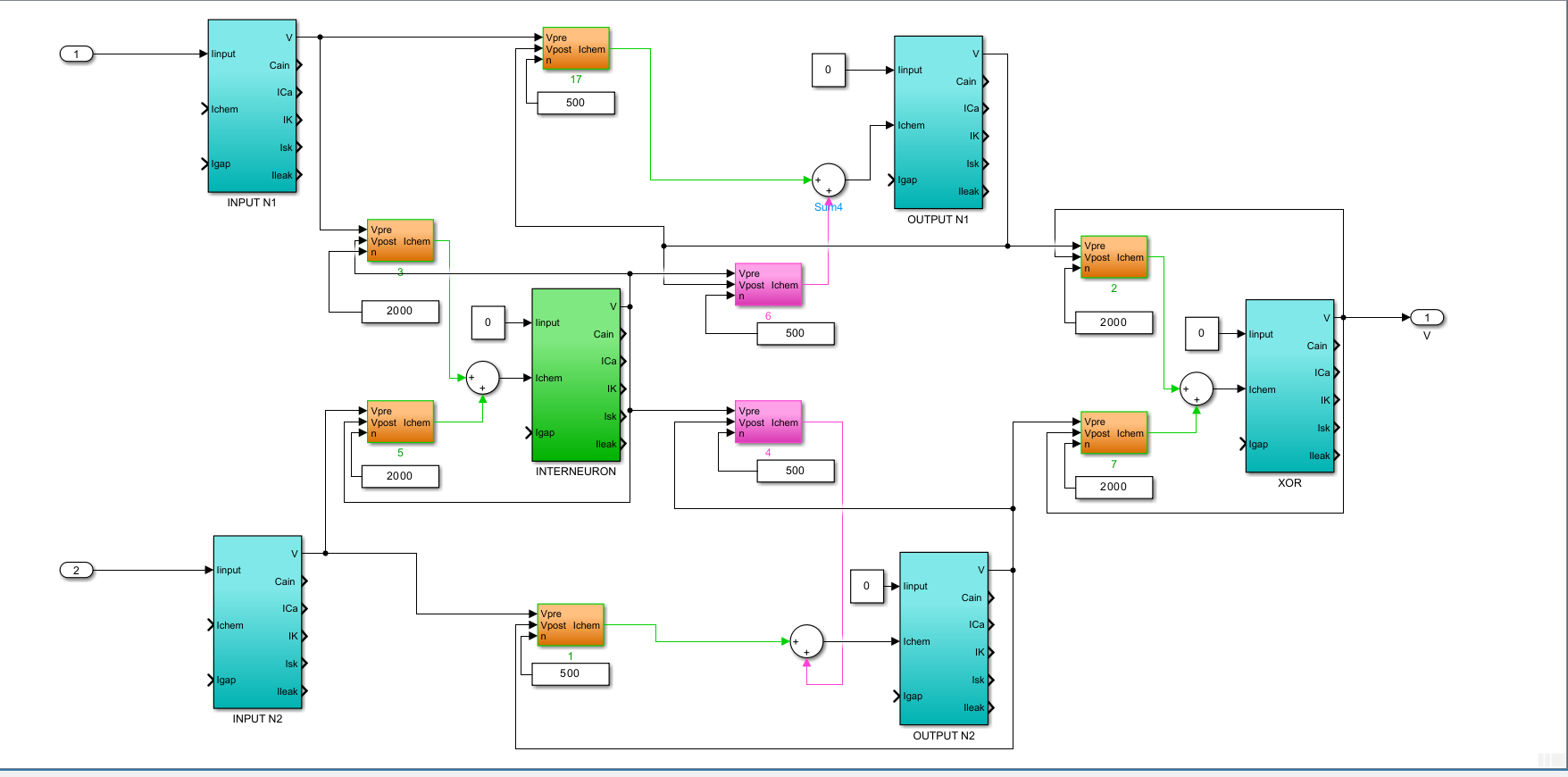} }
	\caption{XOR comparator scheme implemented in SIMULINK using as components the neurons from the simulation of C.Elegans neurons.}
	\label{fig:XOR-Simulink}
\end{figure}

We have tested how the switch operates as expected when it asynchronously processes two signals, with similar amplitudes, an example is shown in Figure 3.

\begin{figure}[h!]
	\centering
       \adjustbox{cfbox=black 1pt}{ 
      \includegraphics[width=0.9\textwidth]{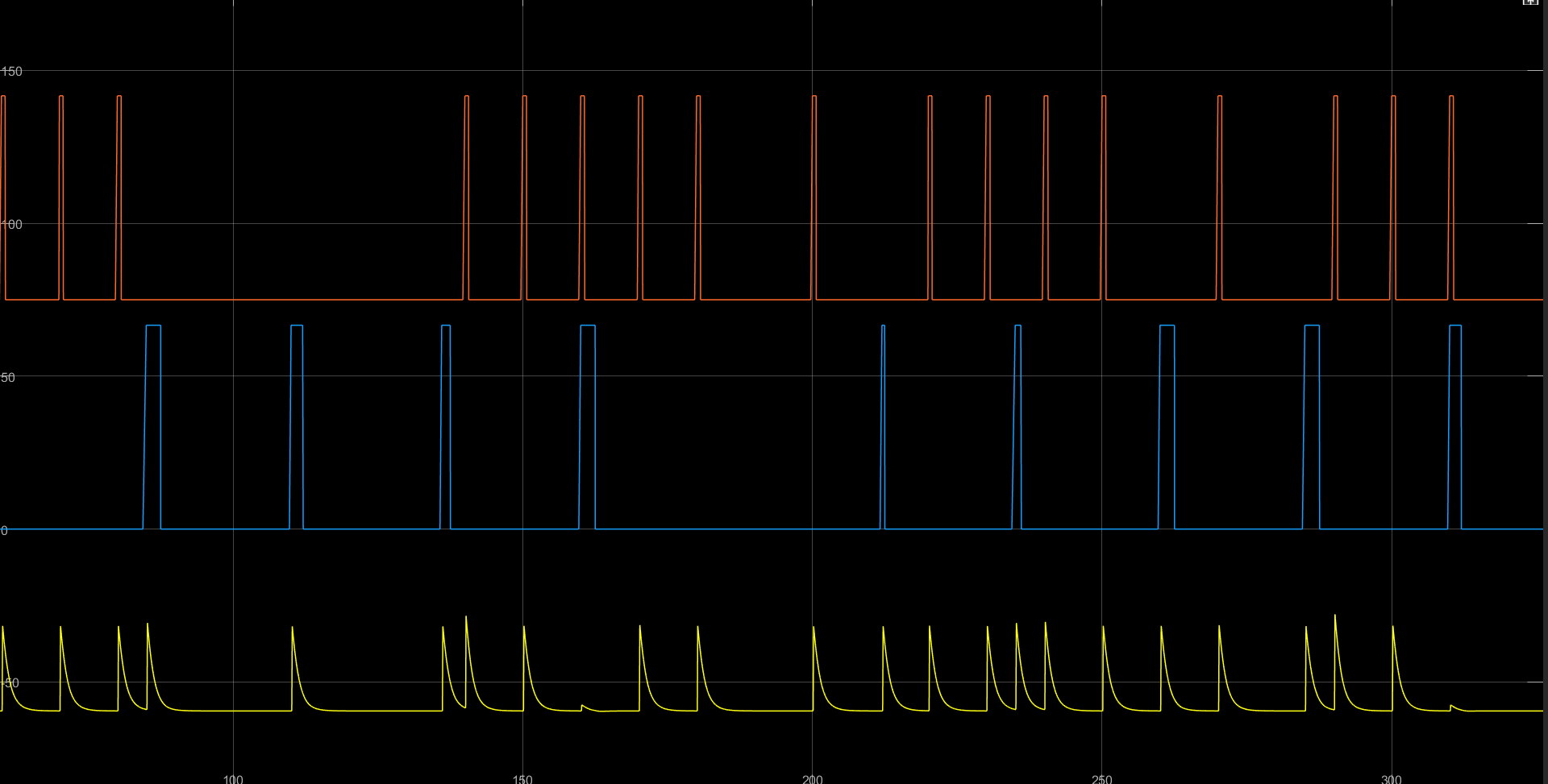}}
     \caption{Example of XOR circuit voltage output (yellow line) with two input signal pulses (red and blue lines). Notice that the coincidence of both input signals results in a null output, as expected.}
	\label{fig:XOR-example}
\end{figure}

\section{Feasibility of biological XOR motifs}

The proposed XOR motif is in fact a simple extension of the well-known lateral inhibition motif, one of the basic core circuit motifs (Luo, 2021).  As expected, the XOR motif is present in the C.Elegans connectome, and we have also checked the evolution along its development using the data from the Worm wiring repository (WormWiring)
Table 1 below presents the result of a preliminary check of the presence of the XOR motif as a directed graph in the global connectome graph.

\begin{table}[h!]
\centering
\begin{tabular}{lcccc}
\toprule
\textbf{Dataset} & \makecell{\textbf{\# chemical} \\ \textbf{synapses}} & \makecell{\textbf{\# instances} \\ \textbf{XOR motif}} & \makecell{\textbf{\# chemical synapses} \\ \textbf{+ gap junctions}} & \makecell{\textbf{\# instances XOR motif} \\ \textbf{including gap junctions}} \\
\midrule
1L1  & 775  & 11  & 858  & 6   \\
2L1  & 986  & 16  & 1110 & 35  \\
3L1  & 1012 & 25  & 1106 & 30  \\
4L1  & 1136 & 41  & 1345 & 105 \\
5L2  & 1515 & 80  & 1807 & 376 \\
6L3  & 1525 & 72  & 1739 & 228 \\
7A   & 2202 & 209 & 2493 & 478 \\
8A   & 2186 & 285 & 2496 & 347 \\
\bottomrule
\end{tabular}
\\
\caption{Abundance of the XOR motif in the C. Elegans connectome along its developmental stages, 1L1 to 8A (adult)}
\label{tab:celegans}
\end{table}

However, the real feasibility of the XOR motif requires a more detailed analysis of specific neuronal circuits, in particular on the type of neurons involved, and the strength and polarity of the connections established (Rakowski et al., 2013). The table above is significant as it shows the popularity of this motif, that is implemented as a DAG (directed acyclic graph) with 6 nodes and 8 edges, as shown in figure 1. 

\section{Plasticity and learning in the XOR circuit}

We have set up this XOR circuit using the detailed neuron model scheme provided in (Hasani 2017), and have established the connection strengths through a heuristic estimation. Parameter guessing is a well-known approach in machine learning for this type of circuits, and in particular for implementing an XOR function, but what could be the origin of this motif in biological systems?

Neurons, as other cells, have an evolutionary story, and as long as their internal model is realistic, we do not need additional arguments. The scheme of the connections is also feasible, given the intrinsic complexity observed in the connectomes even of simplest organisms, like it is the case for C.Elegans. However, we may doubt that the specific, although not unique, strengths used for the synaptic connections are natural. Here there are two possible answers. 

The first one is again evolutionary: if the XOR neural motif has evolved into such configuration due to mutations, and it provides a useful functional response (the comparison of two neuronal signals), it may provide an advantage for natural selection, and get genetically imprinted. 

The second possible answer is trickier: the strengths of the connections could be the result of a plasticity-based learning process over a basic motif involving these six neurons, initially joined though very weak synaptic connections, that would get reinforced with each new signal pulse, up to a saturation/consolidation value. This could be considered as a very basic plasticity mechanism, as there is no feedback to control the evolution, but it is interesting as a potential answer to the origin of the XOR switch without the need for evolutionary arguments.  We have checked that under the configuration proposed, and starting with very low values of the weights, the system saturates the links independently of the type of input signals received, as expected. We have also incorporated the habituation effect, decreasing the strength of the connection for a repeated set of pulses. 

The result is shown in figure 4: the XOR circuit after a brief initial transition period, tends to saturate the connections and operate as expected. 

\begin{figure}[h!]
      \centering
       \adjustbox{cfbox=black 1pt}{ 
      \includegraphics[width=0.8\textwidth]{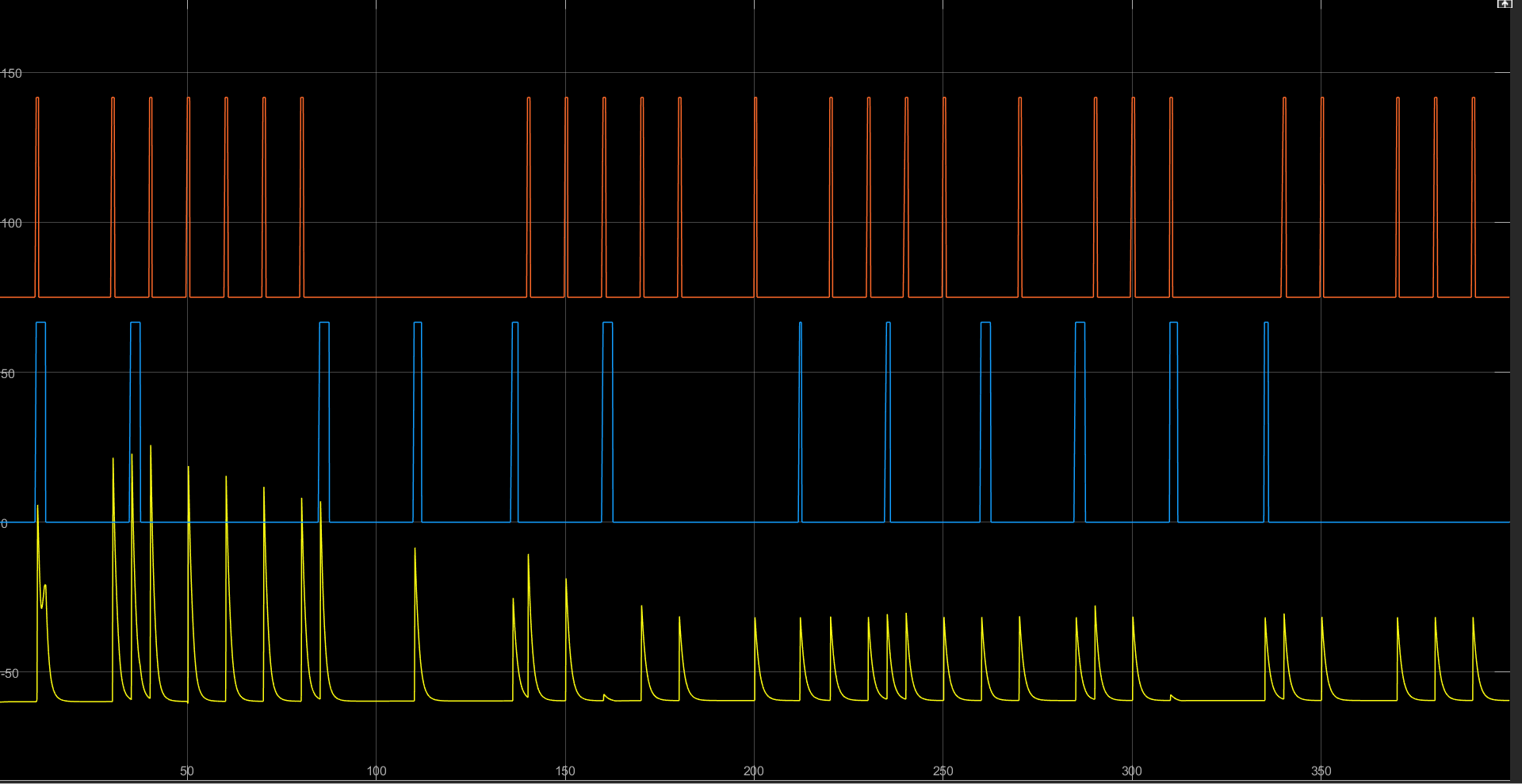}}
      \caption{Example of XOR circuit as in figure 3, in this case the synaptic connections strengths started at a minimum value and were reinforced with each new pulse, until saturation. The saturation values were chosen to be those used in the previous basic XOR model.}
	\label{fig:XOR-Reinforce}
\end{figure}

To test the plasticity, or expressivity, of this simple neural XOR motif, we have implemented it using a computational recurrent neural network. We have selected the framework provided by the Neural Circuit Policies (NCP) to implement Liquid Time Constant Neural Networks (Lechner et al., 2018), as it has derived from the studies previously cited on the properties of the C.Elegans connectome. 
Figure 5 shows the scheme implemented, and we have trained it to learn an arbitrary binary pattern implementing the XOR function.

\begin{figure}[h!]
	\centering
       \adjustbox{cfbox=black 1pt}{ 
	\includegraphics[width=0.6\textwidth]{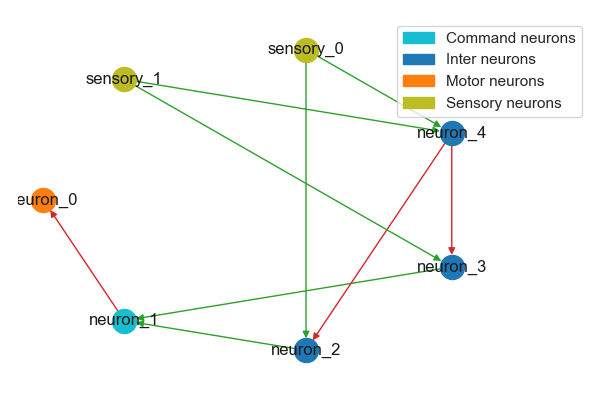}}
	\caption{Graph showing the implementation of an XOR motif based on LTC and using the NCP framework.}
	\label{fig:XOR-NCP}
\end{figure}

We have trained this neural network using 96 bit data sequences (see figure 6) to learn the XOR function using a typical classification loss function (BCEwithLogit loss) and found it converging reasonably after less than 500 iterations, and predicting correctly any random XOR sequence. 

\begin{figure}[h!]
	\centering
       \adjustbox{cfbox=black 1pt}{ 
       \includegraphics[width=0.45\textwidth]{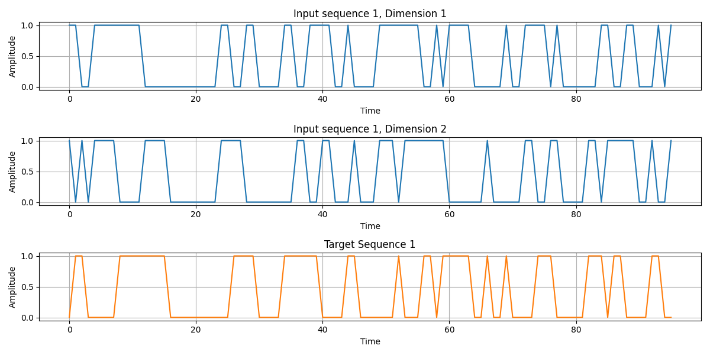}
       \includegraphics[width=0.45\textwidth]{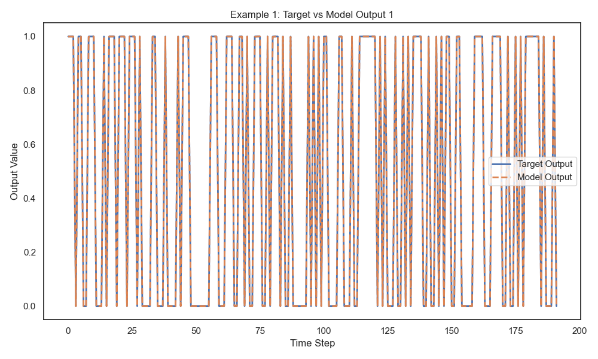}}
	\caption{Left: Binary sequences (96 bits) used to train the XOR LTC neural network. Right: Prediction and target sequences for the XOR LTC neural network on a 192 random binary sequence.}
	\label{fig:binary-seq}
\end{figure}

This exercise shows that the plasticity of this set of neurons conforming the motif is enough to provide an XOR function.

\section{Towards plasticity and learning with XOR feedback}
The following question is how this XOR motif can be integrated into a more complex neuronal circuit to implement a basic circuit for learning.
The basic idea is that the XOR output must provide feedback to the rest of the neuronal circuit in such a way that it is able to “learn” signal sequences. Here, ideally, the word “learn” could mean that the circuit is able to recognize a given signal, store it, classify it and recover it when required. Following our ideas from the previous paper, we will explore how this process may happen using a basic autoencoder.
We consider an autoencoder with a triple binary input, an encoder circuit, a decoder circuit and a triple binary output. Along the training process, the embedded XOR motif should “compare” the binary inputs and the binary outputs, providing the required feedback to adapt the synaptic connection weights of the neural network to converge.  
An scheme as simple as possible can be derived from the XOR motif itself, and it is shown in figure 7.

\begin{figure}[h!]
	\centering       
       \adjustbox{cfbox=black 1pt}{ 
       \includegraphics[width=0.6\textwidth]{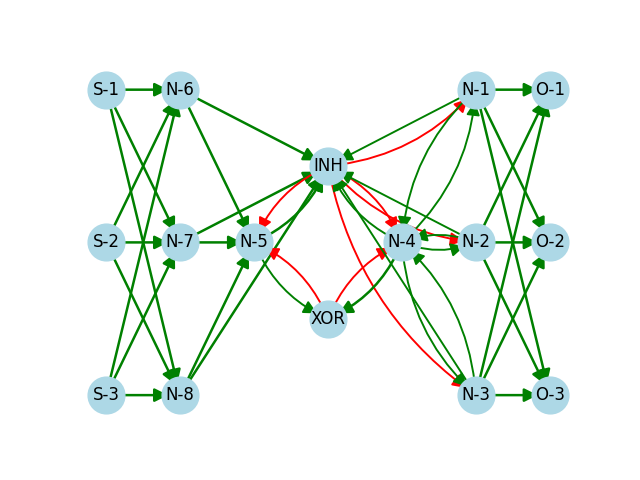}}
	\caption{Scheme of the MULTIXOR neuronal circuit where the core XOR motif is implemented with 2 inhibitory (INH, XOR) and 2 excitatory (N-6, N-9) neurons.}
	\label{fig:MultiXOR}
\end{figure}

The idea behind this MULTIXOR motif is the following: the inhibitory feedback from the XOR neuron via N-6 and N-9 modifies the weights of the connections towards a null value for the XOR. Although this scheme may seem arbitrary, these feedback connections implement in fact an analogy to a basic gradient descent.
This configuration seems biologically feasible, it is a sparse graph, with feedback provided mainly using inhibitory neurons. Notice that the circuit can be “redrawn” to make it closer graphically to what one expects to find when observing a layered structure in the neocortex.
To test the plasticity/expressivity of this configuration as a computational neural network, we have again used the LTC/NCP framework. The corresponding LTC neuronal circuit, shown in figure 8, was trained to auto encode different sequences of length 24x3 bits, that were derived from five popular melodies that use only three different musical notes, reproducing all of them exactly as shown in figure 9.

\begin{figure}[h!]
	\centering       
       \adjustbox{cfbox=black 1pt}{ 
       \includegraphics[width=0.7\textwidth]{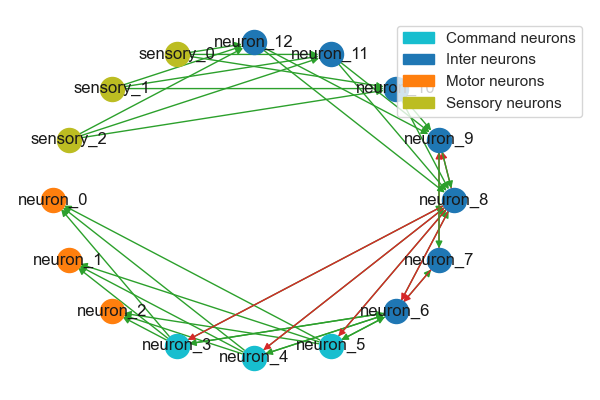}}
	\caption{Graph showing the implementation of the MULTIXOR motif based on LTC and using the NCP framework.}
	\label{fig:NCP-MultiXOR}
\end{figure}

\begin{figure}[h!]
	\centering
       \adjustbox{cfbox=black 1pt}{ 
       \includegraphics[width=0.8\textwidth]{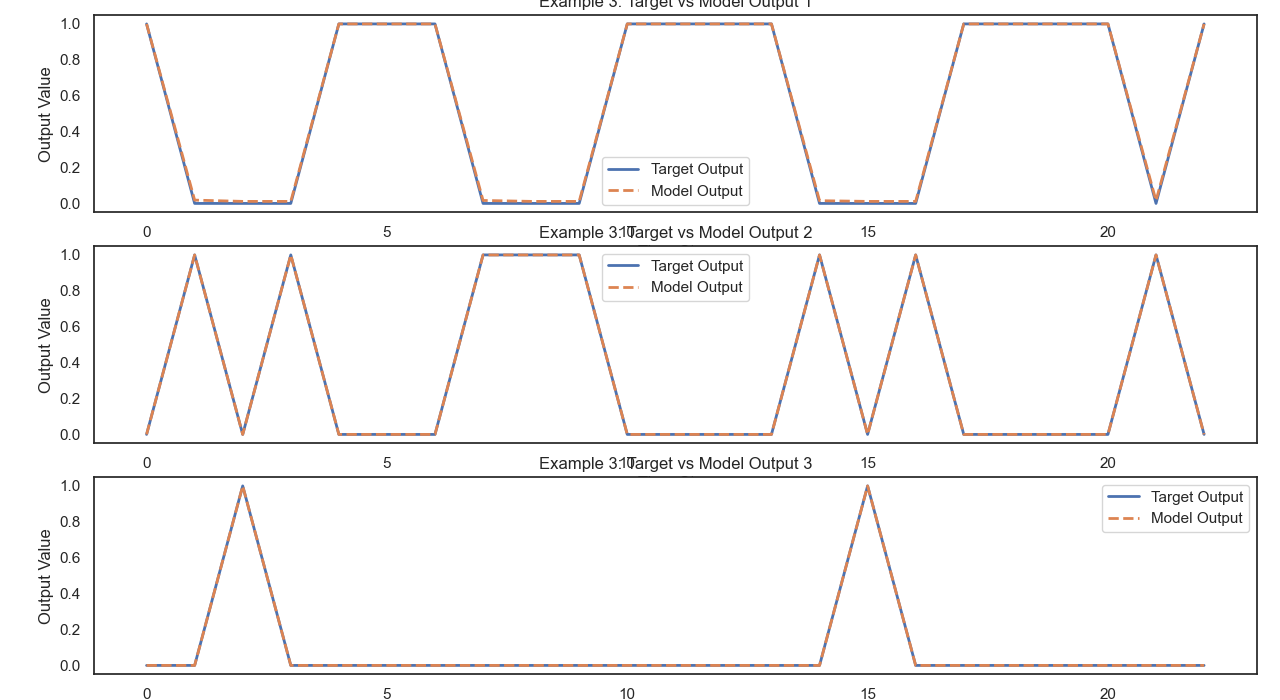}}
	\caption{Example of the binary values in the MULTIXOR autoencoder input-output for the three notes of a melody.}
	\label{fig:Seq-MULTIXOR}
\end{figure}

The previous example is a basic one, and not a very realistic one: typically each neuron will have many more connections, o(100), through its axon and dendrites, and such a simple motif will be part of a much more complex sparse neural network.
However, with a simple block as the XOR motif, we can build much more complex structures to process binary sequences, for example an inverter. In fact, the same motif is able, if adequately trained/configured, to produce the other three basic Boolean operators patterns, AND, OR, NAND, as we have also tested using the same LTC/NCP configuration. So, potentially all type of circuits for signal processing could be built using this simple motif.
This possibility to implement logic circuits using biological neurons is not new, (Koch, Poggio and Torre, 1983),(Fromherz and Gaede, 1993),(Yoder, 2009),(Zeigler and Muzy, 2017), and opens up the possibility to define basic memory units, like D-latches, with or without external regulation, and also circuits implementing logical causality.

\section{Related work}

We do not aim to provide a complete review of the literature discussing these interesting topics, but we would like to point briefly to some of the work that we have considered.
Simple neuronal motifs describing circuits involved in learning in our brain can be found in many recent references based on the study of the connectome using electron microscopy, and we were inspired particularly by those presented in (Campagnola et al., 2022).
A key work to develop this proposal is the detailed neuronal model for the C.Elegans and its implementation in the SIMULINK platform presented in (Hasani 2017), We thank the authors for providing the open versions in github, as it has been very helpful when exploring our ideas.
The work started by this team studying the C.Elegans has evolved into a very nice model for recurrent neural networks, the Liquid Time Constant neuronal networks ((Hasani et al., 2021), much more powerful computationally than the ideas shown in this preprint, but requiring a purely computational training including a mathematical loss function, and a minimization algorithm, and we have used the open version available in github (Lechner et al., 2018) to explore the plasticity of the configurations proposed, although our approach using the XOR circuit to compare two signals and provide feedback to control learning implies a different biological approach.   
The idea of implementing an XOR circuit using a recurrent neuronal network is not new: a similar scheme can be found for example in (Nikolić, 2023), although in our proposal the role of the inhibitory neurons is very clear, and the idea of homeostasis as the key output is new, and it could provide a hint to the observed ratios E/I.
In fact, as stated previously, the XOR motif can be reinterpreted as a basic example of a lateral inhibition circuit, and it is likely the case that parallel XOR circuits, like the MULTIXOR circuit presented here, provide a very similar architecture.

\section{Conclusions and future work}

We have proposed in this paper a very simple motif, a biological neuronal XOR switch, as a key component to provide a comparative useful feedback to control the learning process in a simple neural network. The main objective of the paper is to open the possibility of learning by comparing two signals, and providing feedback for a recurrent training, without the need of other complex mechanism, and without an explicit “reason” to learn but the maintenance of the homeostasis along the processing of the incoming signal, a very basic biological argument.  
An important point is the comparison with a usual RNN, where the neural network takes as input a time-ordered sequence of input data, of limited length. Such computational networks in simple cases may achieve a remarkably fast convergence, and this inspired us to consider that learning is feasible in a short training time for a similar architecture.
The extension of the scheme proposed to handle a typical sensory signal of length limited to 1 second in real time, requires at least o(100-1000) input channels, and such discussion is left for another paper in preparation about processing of melodies as sound signals.
We also want to continue exploring simple motifs built using excitatory and inhibitory neurons, and in particular the implementation of memory latches and of causal rules, that could be useful at a higher level of processing the information.
It would be great if we could test within the connectomes already published, the presence and ratios of motifs proposed in this preprint, as it has been the case for other, slightly simpler, motifs with three and four nodes (Sporns and Kötter, 2004). And it would be ideal to contrast how any of these motifs, if found, performs dynamically, in particular the details of the plasticity mechanism.
As stated in the introduction, this paper tries to address at a very basic, biological level, some of the key points posed in a previous preprint, in particular the possibility that our brain encodes the information that it receives, processes the information and learns it, following a scheme similar to what we use in our encoders in machine learning. 
Our next paper will address, on the other direction of abstraction, how adequate structures for the latent space in our neuronal networks, could support multimodal memory and assemble episodic memory, extending the ideas already presented about concept neurons.

\end{document}